\documentclass{article}
\usepackage[greek.polutoniko,english]{babel}
\usepackage[utf8]{inputenc} 
\usepackage[a4paper, total={6in,8in}]{geometry}
\usepackage[affil-it]{authblk}
\usepackage{amsmath}
\usepackage{amssymb}
\usepackage{mathtools}
\usepackage{graphicx}
\usepackage{subcaption}
\usepackage{booktabs}
\usepackage{setspace}

\large
\begin{document}

\title{A commented translation of Boltzmann’s work, ``Ueber die sogenannte \textit{H}-Curve."}
\author[1,2,3]{Jae Wan Shim}

\affil[1]{Extreme Materials Research Center, Korea Institute of Science and Technology,
            {5 Hwarang-ro 14-gil, Seongbuk}, 
            {Seoul},
            {02792}, 
            {Republic of Korea}}

\affil[2]{Climate and Environmental Research Institute, Korea Institute of Science and Technology,
            {5 Hwarang-ro 14-gil, Seongbuk}, 
            {Seoul},
            {02792}, 
            {Republic of Korea}}
            
\affil[3]{{Division of AI-Robotics, KIST Campus, University of Science and Technology},           
            {5 Hwarang-ro 14-gil, Seongbuk}, 
            {Seoul},
            {02792}, 
            {Republic of Korea}}
\date{}
\maketitle

\begin{abstract}
Boltzmann’s work, ``Ueber die sogenannte \textit{H}-Curve," discusses his demonstration of the essential characteristics of the \textit{H}-curve in a clear, concise, and precise style, showcasing his efforts to persuade his peers. To make these findings more widely accessible, the author aims to provide a translated version of the original article, while also correcting some typographical errors in the mathematical expressions with explanatory footnotes. The final section offers concluding remarks with graphs and relevant references for interested readers.
\end{abstract}


\section{Introduction}
\label{sec:intro}
After conducting a thorough examination of Boltzmann's papers, I have come across a particular article \cite{boltzmann1898} for which no translation was readily available. In this article, Boltzmann adeptly demonstrates the important properties of the \textit{H}-curve with a clear, concise, and precise style, showcasing his efforts to persuade his peers. I have identified some typographical errors in the original mathematical expressions, providing footnotes to correct these errors.

Despite the essential role played by Boltzmann entropy in expanding our comprehension of entropy and its diverse applications across various scientific domains, this concept has encountered substantial challenges and criticisms, first through the Loschmidt\cite{boltzmann1877replyLoschmidt, boltzmann1970incollection, thomson1875} and Zermelo\cite{zermelo1896a, boltzmann1896a, zermelo1896b, boltzmann1897b}\footnote{Author's note: \textit{Annalen der Physik} has been referred to by the name of its editors, such as \textit{Wiedemanns Annalen} during Gustav Heinrich Wiedemann's tenure as editor-in-chief from 1877 to 1899.} paradoxes.

Loschmidt pointed out that if the laws of physics are time-reversible, as Boltzmann assumed, then the increase in entropy over time should be reversible as well. This seemed to contradict Boltzmann's theory that entropy should always increase over time. And Zermelo questioned how to harmonize Boltzmann's statistical perspective on entropy with the observation that physical systems tend to move toward equilibrium, where entropy is at its highest. Zermelo pointed out that a system of ideal gas particles eventually goes back to their initial positions and speeds under strict Hamiltonian evolution, effectively recreating the gas's original state. This phenomenon is called Poincaré recurrence.

The so-called \textit{H}-curve is a plot of the value of \textit{H} over time, where \textit{H} corresponds to entropy with a constant multiplication factor. This curve illustrates the characteristics of the time evolution of entropy. At the end of the original paper by Boltzmann, he wrote ``Vienna, Christmas 1897," and the paper was published the following year in 1898. This paper was published over 25 years after Boltzmann first introduced the concept of statistical interpretation of entropy, and it serves as one of the testaments to how he overcame the objections and disputes surrounding his theory of statistical entropy interpretation.

The forthcoming section presents a translated version of the Boltzmann's paper\cite{boltzmann1898} for the reader's convenience. The paper highlights important properties of the \textit{H}-curve, and provides insights into Boltzmann's attempts to persuade his contemporaries. In the final section, I offer concluding remarks with graphs and provide references to related works that may be of interest to the readers.

\section{On the so-called \textit{H}-Curve}
\label{sec:H-Curve}
\subsection{Translator's Note}
This translation endeavors to preserve both the style and sentence structure inherent in Boltzmann's original paper, a reflection of the writing norms from over a century ago. This commitment extends beyond the scientific vernacular of the time, encompassing broader linguistic nuances as well. Although the translation strives to maintain accuracy to the original text, readers may find certain expressions to be unconventional or antiquated compared to contemporary writing styles and terminology.
\subsection{Translation}
Allow me to briefly discuss the properties of the curve used for the illustration of certain propositions in the theory of gases\footnote{Boltzmann's note: Nature 51, p. 413, Feb. 28, 1895. Again, Ann. Vol. 60, p. 392, 1897. Therein, even a figure of an \textit{H}-curve is found.}, here independently of their significance in the theory of heat. These properties emerge most clearly and in the simplest manner when the construction of the curve is linked to a very trivial example of probability theory.

Let us first consider the following simple case: In an urn, there are an equal number of white and black balls, otherwise of completely identical characteristics. Guided by chance, we make an arbitrary odd number (2${N}$+1) of draws from the urn, which we denote in sequence as
$$Z_{-N}, Z_{-N+1}, \ldots, Z_0, Z_1, Z_2, \ldots, Z_N.$$
After each draw, we return the drawn ball back to the urn.

Let us also assume that $n$ is any even integer that is smaller than $2N+2$. We denote by $a_k$ the total number of white balls that were drawn during the $n$ draws $Z_{k}, Z_{k+1}, \ldots, Z_{k+n-1}$, where $k$ can be any of the $2N-n+2$ positive or negative integers including zero, which lie between $-N$ and $N-n+1$, inclusive.

We now construct a rectangular coordinate system in the plane. To every value of $k$, there corresponds a point $B_k$ whose abscissa\footnote{Author's note: The abscissa refers to the horizontal (x) axis, while the ordinate refers to the vertical (y) axis.} is $$O A_k = x = k/n$$ and whose ordinate $A_k B_k = y$ is equal to the always positively taken numerical value of the expression $1-2a_k/n$. If we denote this numerical value\footnote{Boltzmann's note: One could also set $y= (1-2a_k/n )^b$ and understand $b$ to be the number $2$ or another positive number. One then obtains a greater variety of \textit{H}-curves. However, as I am not at all concerned with an exhaustive geometric discussion of all possible structures related to the \textit{H}-curve but only with the most concise possible visualization of the properties of the same found in physics, I limit the discussion to the case considered in the text.} with a bar placed above it, then
\begin{equation}
y= \overline {1-\frac {2a_k}{n} }.
\end{equation}

We thus obtain a series of $2N - n + 2$ discrete points $B_{-N}, B_{-N+1}, \ldots, B_{N+1-n}$. We now require the number $n$ and the number $N$ to be very large compared to $1$, but the latter number to an even higher degree, so that the quotient $N/n$ is also very large compared to $1$, while we are indeed thinking of arbitrarily large but always finite (not metaphysically infinite) numbers. Then, any two adjacent points $B_k$ and $B_{k+1}$ get very close to each other, as the difference of their abscissae as well as their ordinates vanishes with increasing $n$; the former is equal to $1/n$, the latter has at most the value $2/n$.

We will therefore call the series of $B$ points a curve, specifically the \textit{H}-curve of the lottery, without intending to claim that it actually possesses all the properties that are otherwise demanded of curves in geometry. Namely, this curve lacks the property that the position of each point is defined by an unchangeable analytical formula. Rather, this position is determined by the result of a real process dependent on unknown causes, one could almost say left to chance. Nevertheless, it cannot be denied that if $N$ is any arbitrarily large number, one can really make $2N+1$ draws in the described manner and, if one also assumes a specific value for $n$ that is small compared to $N$ but large compared to $1$, one could draw the corresponding \textit{H}-curve \textit{i.e.}, the corresponding $2N-n+1$ points\footnote{Author's note: The number of points should be $2N-n+2$.} $B$. If one were to make another $2N + 1$ draws, and then yet another $2N + 1$ draws, one could draw any number of other \textit{H}-curves, which would surely not be identical to each other but would nonetheless have certain common characteristics, the discovery of which is precisely the matter at hand.

Since $n$ is assumed to be very large, $a_k$ is most likely, \textit{i.e.}, for the vast majority of the values of $k$, very close to $n/2$, hence $y$ is very close to zero. The  \textit{H}-curve therefore almost coincides with the abscissa axis at almost all points. However, if one chooses $N$ large enough, one also obtains points on the \textit{H}-curve where it deviates by a finite amount from the abscissa axis. We call such points humps.

If we set, for example, $N = 1000 \cdot 2^n$, where $n$ is already a very large number, then we have a chance that over the entire course of all $2N + 1$ draws from $Z_{-N}$ to $Z_N$, there are $2000$ instances where a black ball is drawn in a series of $n$ consecutive draws. If $k$ is the index of the first draw in such a series, then $a_k = 0$, hence $A_k B_k=y=1$. Similarly, the ordinate of the  \textit{H}-curve is equal to $1$ when a white ball is drawn $n$ times in succession, which for $N=1000\cdot 2^n$ also occurs approximately $2000$ times over all $2N+1$ draws. The largest possible hump, whose highest ordinate is equal to $1$, therefore occurs about $4000$ times over all $2N + 1$ draws. The number of humps of lesser but finite height different from zero is much larger.

Despite its continuity, the \textit{H}-curve considered does not possess a tangent in the strictest sense of the word, \textit{i.e.}, the direction of the chord drawn from a specific point on the curve to a very nearby point does not necessarily approach a certain limit as the latter point increasingly nears the former.

For instance, let $k$ increase by one unit, thus transitioning from point $B_k$ to point $B_{k+1}$, then the abscissa increases by $1/n$. Also, $a_k$ is the number of white balls drawn in the draws $Z_k, Z_{k+1},\ldots, Z_{k+n-1}$. If we let $k$ increase by one unit, the draw $Z_k$ is excluded, but the draw $Z_{k+n}$ is added. If the same color was drawn in both of these draws, the value of $a_k$, and therefore also of $y$, does not change. The line $B_k B_{k+1}$ is therefore parallel to the abscissa axis. In contrast, if balls of different colors were drawn, the value of $a_k$ changes by one unit. The change in the ordinate when transitioning from point $B_k$ to point $B_{k+1}$ thus has the numerical value $2/n$ according to equation $(1)$, and is twice as large as the increase in the abscissa, so that the line $B_k B_{k+1}$ forms an angle with the abscissa axis on one side or the other, the tangent of which is $2$.

Nevertheless, in the majority of cases, the chord drawn from point $B_k$ to point $B_{k+l}$ approaches a limit, which we call the quasitangent, when $l$ is made large compared to one unit but smaller than $n$.

For example, let the ordinate of $B_k=1$, so that $B_k$ is the point with the highest ordinate of a hump of the highest possible height. Then the corresponding $a_k$ is either zero or $n$. We consider the first case $a_k = 0$, \textit{i.e.}, no white ball was drawn in the draws $Z_k, Z_{k+1},\ldots, Z_{k+n-1}$. If we move to the point $B_{k+l}$, \textit{i.e.}, if we let $k$ increase by $l$, then the $l$ draws $Z_k, Z_{k+1}, \ldots, Z_{k+l-1}$ are to be omitted, whereas the $l$ draws $Z_{k+n}, \ldots, Z_{k+n+l-1}$ are to be included. We know that in the first $l$ draws, not a single white ball was drawn; in the latter $l$ draws, if $l$ is a large number, nearly $l/2$ white balls probably have been drawn. Therefore, $a_{k+l}$ is likely to be about $l/2$ greater than $a_k$. According to formula (1), the ordinate of the point $B_{k+l}$ is therefore smaller by $l/n$ than that of the point $B_k$, and since the abscissa of the former point is larger by the same amount than that of the latter, the line $B_k B_{k+l}$ is inclined at an angle of $45^\circ$ to the abscissa axis. In a similar way, the direction of the quasitangent can also be calculated when the ordinate of $B_k$ has another finite value different from zero. If $B_k$ has an ordinate only very slightly different from zero, \textit{i.e.}, it belongs to that majority of points on the curve that are very close to the abscissa axis, then the quasitangent is parallel to the abscissa axis.

I may not be mistaken in believing that professional geometers will scoff at the \textit{H}-curve. In reply I would like to recall that the curves drawn by meteorographs, barometer graphs, thermometer graphs, \textit{etc.}, display a character that reminds one of the properties of the \textit{H}-curve. It is not at all ruled out that, in the case of these curves, at any given moment the position of the expected point is determined by a high number of competing causes. These causes, following the laws of probability, result in points sometimes higher sometimes lower, but over a long time they exhibit regularities. Thus, the line drawn towards a somewhat distant point from a given point displays a fairly fixed inclination against the abscissa axis, despite the countless small notches in the curve. Indeed, it may even occur more often than we think, that a force which seems constant to us, is only so on average over a longer interval. If we were able to observe the course of this force in the smallest time scale, it would exhibit fluctuations following the laws of probability.

I intend to further elaborate on one specific property of the \textit{H}-curve. It is evidently irrelevant whether we arrange the $2N+1$ draws in the order $Z_{-N}, Z_{-N+1},\ldots, Z_N$ or in the reversed $Z_{N}, Z_{N-1},\ldots, Z_{-N}$. It is not required that every single \textit{H}-curve be perfectly symmetrical with respect to the ordinate axis, but on average every \textit{H}-curve follows the same laws in the direction of increasing abscissae as it does in the direction of decreasing. It should not be possible to prove that the curve possesses a property for increasing abscissae but not equally possess this property for decreasing ones.

We now wish to plot a curve $J$ that has the same symmetry with respect to the positive and negative abscissae but has a tangent in the ordinary sense at every point. Like the \textit{H}-curve, the $J$ curve should mostly run very close to the abscissa axis; only at particular points should it rise finitely above it. Let us plot all points $P$ of the curve $J$ that are assigned a given unusually large ordinate $y_1$. If we move from any of these points by a finite amount in the positive or negative direction of the abscissa, we will typically reach points with smaller ordinates, but the differential quotient $dy/dx$ is positive for as many of the points $P$ as it is negative. The average value of this differential quotient for all points $P$ is zero.

However, this does not apply to the \textit{H}-curve. Let us again record on this curve all points $Q$ that have an ordinate $y_1$ finitely differing from zero. If we let the abscissa increase by $\Delta x = 1/n$ for all these points, there will be sometimes positive and sometimes negative increases in the ordinate corresponding to these increases in $x$. In contrast, if we set the increase $\Delta x$ of the abscissa equal to $l/n$ for the same points, where $l$ is a large number, then the corresponding increase $\Delta y$ of the ordinate is not only almost exclusively negative, but the average value of the quotient $\Delta y / \Delta x$ for all points Q also approaches a well-defined limit as $l$ continues to grow, as long as it remains smaller than $n$.

Of course, the same applies when progressing in the positive and negative abscissa direction. The limit of the quotient $\Delta y / \Delta x$ is thus equal in magnitude, but with opposite signs for positive and negative $\Delta x$. 

Three cases are possible for the location of the point $Q$. First: The point can lie at the highest peak of a hump of the \textit{H}-curve, so that the curve falls off on both sides. Second: The point can lie so close to the highest spike of a hump that the curve may still rise on one or the other side, or even on both sides, if you increase the abscissa by $1/n$ or a small multiple thereof, but as soon as you let the abscissa grow or decrease by $l/n$, where $l$ is a very large number that can still be small compared to $n$, the ordinate decreases. The third case is that the hump on which the point $Q$ is located, is, at a finite distance, equally high as $y_1$ or exceeds $y_1$ by a finite amount. Now, it is the characteristic property of the \textit{H}-curve that the frequency of its humps decreases so rapidly with increasing height that only a vanishingly small number of points $Q$ occur for which the third case applies.

The described characteristic properties of the \textit{H}-curve are not compromised if one connects every two neighboring points denoted by $B$ with an arbitrary very small curve. Algebraically, such small connecting curves can be constructed in the following way. We assume that the draw $Z_0$ occurs at time zero, the draw $Z_1$ at time $\tau = 1/n$, $Z_2$ at time $2 \tau = 2/n$, and so on up to $Z_k$ at time $k\tau$. Furthermore, we understand $f_k(t)$ to be a function which always has the value zero if a black ball was drawn in the draw $Z_k$. If, on the other hand, a white ball was drawn in the draw $Z_k$, then it should be
\[
f_k(t)=
\begin{cases}
0 & \text{ for }  t\leq(k-n-\nu)\tau,\\
 \varphi[t-(k-n-\nu)\tau] & \text{ for }  (k-n-\nu)\tau\leq t\leq(k-n)\tau,\\ 
 1 & \text{ for } (k-n)\tau\leq t\leq k\tau, \\
 \varphi[(k+\nu)\tau-t] & \text{ for } k\tau\leq t\leq(k+\nu)\tau,\\
 0 & \text{ for } (k+\nu)\tau\leq t. 
\end{cases}
\]

The simplest case is obtained when we assume $\nu=1$. However, $\nu$ can also be set equal to $2$ or $3$, or even a number that is large compared to $1$, as long as it is small compared to $n$. $\varphi(u)$ can be any function of $u$ that has the value zero for $u=0$ and the value of $1$ for $u=\nu\tau$, and which increases continuously from the former value to the latter. If we set\footnote{Author's note: The last summation index should be $k=N$, so that $$ y = \sum_{k=-N+n}^{k=N} f_k(t)$$.}
$$ y = \sum_{k=-N+n}^{k=n} f_k(t)$$
and plot the values of $t$ on the abscissa axis and the corresponding values of $y$ on the ordinate axis, we obtain a curve that is continuous in the mathematical sense. For $\nu=1$, all ordinates of this curve, which belong to values of $t$ that are whole multiples of $\tau$, coincide exactly with the ordinates of the individual points from which the \textit{H}-curve was previously composed. For other values of $\nu$, they deviate only by an infinitesimal amount.

The new \textit{H}-curve, which we call the continuously drawn \textit{H}-curve of the lottery, has a tangent in the usual sense of the word, even if the increase in the abscissa is smaller than $1/n$. However, if the increase in the abscissa $\Delta x$ is large compared to $1/n$ but still small compared to $1$, the quotient $\Delta y / \Delta x$ approaches another limit, which corresponds to the quasitangent.

Of course, my only intention here is to show that curves with these properties are geometrically constructible and that therefore, there can be no contradiction in attributing analogous properties to the \textit{H}-curve that occurs in the theory of gases consisting of a very large finite number of completely enclosed molecules. For such gases, the quantity \textit{H} which measures the probability or the disorder of a state,\footnote{Author's note: Here, \textit{H} is defined with the sign convention proper to an entropy.} increases with extremely high probability when starting from an ordered state, \textit{i.e.} from a state for which \textit{H} finitely differs from its maximum value. Later, \textit{H} then remains equal to its greatest value for an enormously long time, but after an even longer time, it again assumes a value that is finitely different from its maximum value.

The \textit{H}-curve resembles the one first considered when the collisions last for an infinitesimally short time, as then the value of the quantity \textit{H} only experiences a sudden change at the moment of the collision; however, it resembles the continuously made \textit{H}-curve of the lottery when the collisions last for a short but finite time. The reverse sequence of states of the gas is always also possible. It is therefore also possible that \textit{H} is very close to its maximum value at the beginning and deviates significantly from this value in a relatively short time; however, the task of finding an initial state of all gas molecules (we will call it a critical initial state), which meets the latter condition, is in a certain sense ambiguous. Such an initial state is not determined by the corresponding value of \textit{H} but by the fact that the initial positions and the initial motions of all molecules are adjusted to each other in a certain way.

Of course, real bodies can never behave absolutely like systems consisting of a large finite number of gas molecules. This is already the case because the former can never be completely taken out of contact with all other bodies.  The fact that real bodies often behave approximately like systems consisting of a finite number of gas molecules, which initially have an ordered state for which \textit{H} is essentially different from its maximum value, but never like systems of gas molecules that are initially in a critical state, for which \textit{H} initially still has its maximum value but soon becomes significantly smaller, is explained in the mechanical view of nature by assuming that the initial state of the world corresponds to an ordered state of molecules.

Given that the bodies we experiment with are always taken from this world, the probability that they are initially in an ordered state is very high, and this state, if we keep external influences as far away as possible, always transitions to a disordered one. There is no doubt that a world in which all natural processes take place in reversed chronological order is just as conceivable. However, a person living in this reversed world would not have a different perception than us. Such a person would simply refer to what we call the future as the past and vice versa.

Vienna, Christmas 1897.

\section{Concluding remarks and related literature}
To conclude this translation, I reference literatures that introduce Boltzmann's work in greater detail, such as those by Klein, Brush, Uffink, Höflechner, Cercignani, Lindley, Galavotti, and Darrigol where further references can be found\cite{klein1973, klein1970, brush1999, uffink2007, hoflechner1994, cercignani1998, lindley2001, gallavotti1999, darrigol2018}. Notably, the Ehrenfest urn model, being directly influenced by the Boltzmann's urn model, was developed. Using the Ehrenfest urn model, one can demonstrate that the time required for a system -- divided by a permeable membrane into two containers, with one initially empty and the other initially filled with $N$ particles, independently moving from one container to another with probability $1/N$ -- to return to its initial state is $2^N$\cite{ehrenfest1907, klein1970}. For example, with $N = 1000$, it would take about $3.4 \times 10^{293}$ years, assuming particles change containers at a rate of one per second.

\section*{Acknowledgement}
This work was partially supported by the KIST Institutional Program.

\section*{Appendix}
\begin{figure}[!h]
\centering
\includegraphics[scale=1]{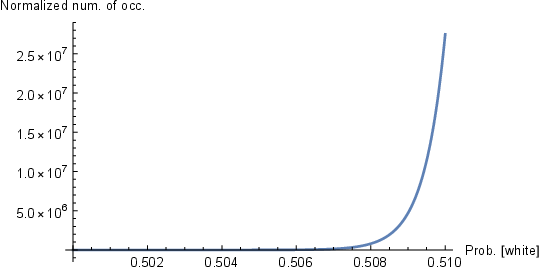}
\caption{Graph of the number of occurrences of the largest possible bump $\mathfrak{N}(p)$ divided by $\mathfrak{N}(1/2)$, as a function of the probability of drawing a white ball $p$. }
\label{fig:fig1}
\end{figure}

\begin{figure}[!h]
\centering
\includegraphics[scale=1]{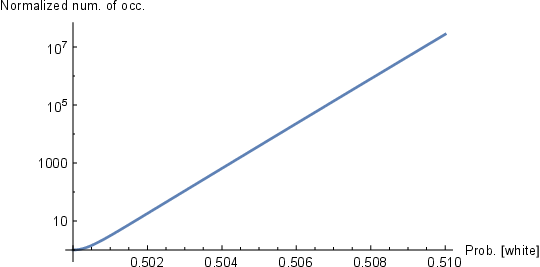}
\caption{The vertical axis of Fig.~\ref{fig:fig1} is presented on a logarithmic scale.}
\label{fig:fig2}
\end{figure}

The urn experiment of Boltzmann, where the urn is filled with white and black balls, can be analyzed in terms of the probability of drawing a white ball $p$ and the number of occurrences of the largest possible bump $\mathfrak{N}(p)$ as a function of $p$. When $p=1/2$, the probability of drawing a white ball is equal to that of drawing a black ball, and the number of occurrences $\mathfrak{N}(1/2)$ is approximately $4000$ times for $N=1000\cdot 2^n$ as demonstrated by Boltzmann. My remark is that when the probability of drawing a white ball is $p=1/2+\varepsilon$, where $\varepsilon$ is a small deviation from $1/2$, the normalized number of occurrences $\tilde{\mathfrak{N}}(p)=\mathfrak{N}(p)/\mathfrak{N}(1/2)$ is approximately given by the formula 
$$\tilde{\mathfrak{N}}(p) \approx 2^{(n-1)} \left[ (p)^n + (1-p)^n\right].$$
This indicates that even a small deviation from $p=1/2$ can result in a significant increase in the number of occurrences of the largest possible bump. For instance, the graph of $\tilde{\mathfrak{N}}(p)$ is shown in Fig.~\ref{fig:fig1} for $n=900$. In Fig.~\ref{fig:fig2}, the vertical axis of Fig.~\ref{fig:fig1} is presented on a logarithmic scale.

\bibliographystyle{unsrt}
\bibliography{commentedTranslationofBoltzmann}

\begin{thebibliography}{10}

\bibitem{boltzmann1898}
Ludwig Boltzmann.
\newblock Ueber die sogenannte \textit{H}-{C}urve.
\newblock {\em Mathematische Annalen}, 50:325--332, 1898.

\bibitem{boltzmann1877replyLoschmidt}
Ludwig Boltzmann.
\newblock {{\"U}ber die Beziehung eines allgemeinen mechanischen Satzes zum
  zweiten Hauptsatze der W{\"a}rmetheorie}.
\newblock {\em Sitzungsberichte der Kaiserlichen Akademie der Wissenschaften.
  Mathematisch-Naturwissenschaftliche Classe.}, 75:67--73, 1877.

\bibitem{boltzmann1970incollection}
Ludwig Boltzmann.
\newblock {{\"U}ber die Beziehung eines allgemeinen mechanischen Satzes zum
  zweiten Hauptsatze der W{\"a}rmetheorie}.
\newblock In {\em Kinetische Theorie II. WTB Wissenschaftliche
  Taschenb{\"u}cher}, volume~67, pages 240--247. Vieweg+Teubner Verlag, 1970.

\bibitem{thomson1875}
William Thomson.
\newblock {The Kinetic Theory of the Dissipation of Energy}.
\newblock {\em Proceedings of the Royal Society of Edinburgh}, 8:325--334,
  1875.

\bibitem{zermelo1896a}
Ernst Zermelo.
\newblock Ueber einen {S}atz der {D}ynamik und die mechanische
  {W}{\"a}rmetheorie.
\newblock {\em Annalen der Physik}, 57:485--494, 1896.

\bibitem{boltzmann1896a}
Ludwig Boltzmann.
\newblock {E}ntgegnung auf die w{\"a}rmetheoretischen {B}etrachtungen des {Hrn.
  E. Zermelo}.
\newblock {\em Annalen der physik}, 57:773--784, 1896.

\bibitem{zermelo1896b}
Ernst Zermelo.
\newblock Ueber mechanische {E}rkl{\"a}rungen irreversibler {V}org{\"a}nge.
\newblock {\em Annalen der Physik}, 59(12):793--801, 1896.

\bibitem{boltzmann1897b}
Ludwig Boltzmann.
\newblock {Zu Hrn. Zermelo's Abhandlung} „{U}eber die mechanische
  {E}rkl{\"a}rung irreversibler {V}org{\"a}nge''.
\newblock {\em Annalen der Physik}, 60:392--398, 1897.

\bibitem{klein1973}
Martin~J. Klein.
\newblock The development of {B}oltzmann's statistical ideas.
\newblock In E.~G.~D. Cohen and W.~Thirring, editors, {\em The Boltzmann
  Equation}, pages 53--106, Vienna, 1973. Springer Vienna.

\bibitem{klein1970}
Martin~J. Klein.
\newblock {\em Paul Ehrenfest. Volume 1. The making of a theoretical
  physicist}.
\newblock Elsevier, 1970.

\bibitem{brush1999}
Stephen~G. Brush.
\newblock Gadflies and geniuses in the history of gas theory.
\newblock {\em Synthese}, 119:11--43, 1999.

\bibitem{uffink2007}
Jos Uffink.
\newblock Compendium of the foundations of classical statistical physics.
\newblock In {\em Philosophy of Physics}, pages 923--1074. Elsevier, 2007.

\bibitem{hoflechner1994}
Walter Höflechner.
\newblock {\em Ludwig Boltzmann: Leben und Briefe}.
\newblock Akademische Druck- u. Verlagsanstalt, 1994.

\bibitem{cercignani1998}
Carlo Cercignani.
\newblock {\em Ludwig Boltzmann: The Man who Trusted Atoms}.
\newblock Oxford University Press, 1998.

\bibitem{lindley2001}
David Lindley.
\newblock {\em Boltzmann's Atom: The great debate that launched a revolution in
  physics}.
\newblock The Free Press, 2001.

\bibitem{gallavotti1999}
Giovanni Gallavotti.
\newblock {\em Statistical Mechanics: A Short Treatise}.
\newblock Theoretical and Mathematical Physics. Springer, 1st edition, 1999.

\bibitem{darrigol2018}
Olivier Darrigol.
\newblock {\em Atoms, Mechanics, and Probability: Ludwig Boltzmann's
  Statistico-Mechanical Writings - An Exegesis}.
\newblock Oxford University Press, 2018.

\bibitem{ehrenfest1907}
Paul Ehrenfest and Tatjana Ehrenfest.
\newblock {\em Physikalische Zeitschrift}, 8:311--314, 1907.

\end{thebibliography}
\end{document}